# Feedback Vertex Set in Mixed Graphs

Paul Bonsma[1]   Daniel Lokshtanov[2]

**Abstract**

A mixed graph is a graph with both directed and undirected edges. We present an algorithm for deciding whether a given mixed graph on $n$ vertices contains a feedback vertex set (FVS) of size at most $k$, in time $2^{O(k)}k!O(n^4)$. This is the first fixed parameter tractable algorithm for FVS that applies to both directed and undirected graphs.

## 1 Introduction

For many algorithmic graph problems, the variant of the problem for directed graphs (*digraphs*) is strictly harder than the one for undirected graphs. In particular, replacing each edge of an undirected graph by two arcs going in opposite directions yields a reduction from undirected to directed graphs for most network design, routing, domination and independence problems including SHORTEST PATH, LONGEST PATH and DOMINATING SET.

The *Feedback Vertex Set* problem is an exception to this pattern. A *feedback vertex set (FVS)* of a (di)graph $G$ is a vertex set $S \subseteq V(G)$ such that $G - S$ contains no cycles. In the *Feedback Vertex Set* (FVS) problem we are given a (di)graph $G$ and an integer $k$ and asked whether $G$ has a feedback vertex set of size at most $k$. Indeed, if one replaces the edges of an undirected graph $G$ by arcs in both directions, then every feedback vertex set of the resulting graph is a *vertex cover* of $G$ and vice versa. Hence, this transformation can not be used to reduce FEEDBACK VERTEX SET in undirected graphs to the same problem in directed graphs, and other simple transformations do not seem possible either. Thus FVS problems on undirected and directed graphs are different problems; one is not a generalization of the other. This is reflected by the fact that the algorithms for the two problems differ significantly across algorithmic paradigms, be it approximation [2, 1, 10], exact exponential time algorithms [13, 14, 23] or parameterized algorithms [3, 7, 5, 6]. *In this paper we bridge the gap between the parameterized algorithms for* FEEDBACK VERTEX SET *by giving one algorithm that works for both.* More generally, we give the first algorithm for FVS in *mixed graphs*, which are graphs that may contain both edges and arcs. Cycles in mixed graphs are defined as expected: these may contain both edges and arcs, but all arcs should be in the same direction (see Section 2 for precise definitions).

For a mixed graph $G$ on $n$ vertices and an integer $k$, our algorithm decides in time $2^{O(k)}k!\,O(n^4)$ whether $G$ contains a FVS $S$ with $|S| \le k$, and if so, returns one. Algorithms of this type are called *Fixed Parameter Tractable (FPT) algorithms*. In general, the input for a *parameterized problem* consists of an instance $X$ and integer parameter $k$. An algorithm for such a problem is an FPT algorithm if its time complexity is bounded by $f(k) \cdot O(|X|^c)$,

[1] Humboldt-Universität zu Berlin, Institut für Informatik,Unter den Linden 6, 10099 Berlin, Germany, email: bonsma@informatik.hu-berlin.de. Supported by DFG project "Fast parametrized algorithms for directed graph problems"

[2] University of California, San Diego, Department of Computer Science and Engineering, 9500 Gilman Drive, La Jolla, CA 92093-0404, USA, email: dlokshtanov@cs.ucsd.edu.

where $|X|$ denotes the input size of $X$, $f(k)$ is an arbitrary computable function of $k$, and $c$ is a constant independent of $k$. The function $f(k)$ is also called the *parameter function* of the complexity, or of the algorithm. Since the first systematic studies on FPT algorithms in the '90s (see e.g. [8]), this has become a very active area in algorithms. See [12, 20] for recent introductions to the area.

FEEDBACK VERTEX SET is one of the classical graph problems and it was one of the first problems to be identified as NP-hard [18]. The problem has found applications in many areas, see e.g. [11, 6] for references, with one of the main applications in *deadlock recovery* in databases and operating systems. Hence the problem has been extensively studied in algorithms [2, 1, 10, 13, 14, 23, 25]. The parameterized complexity of FEEDBACK VERTEX SET on undirected graphs was settled already in 1984 in a monograph by Melhorn [19]. Over the last two decades we have seen a string of improved algorithms [3, 8, 9, 21, 17, 22, 15, 7, 5] (in order of improving parameter function), and the current fastest FPT algorithm for the problem has running time $O\left(3.83^k k n^2\right)$ [4], where $n$ denotes the number of vertices of the input graph. On the other hand, the parameterized complexity of FEEDBACK VERTEX SET on DIRECTED graphs was considered one of the most important open problems in Parameterized Complexity for nearly twenty years, until an FPT algorithm with running time $O\left(n^4 4^k k^3 k!\right)$ was given by Chen et al [6] in 2007. Interestingly, in [16], the permanent deadlock resolution problem as it appears in the development of distributed database systems, is reduced to feedback vertex set in mixed graphs. However, to the best of our knowledge, no algorithm for FVS on mixed graphs has previously been described until now.

We now give an overview of the paper. We start by giving precise definitions in Section 2. In Section 3 we give an sketch of the algorithm, and outline some the obstacles one needs to overcome in order to design an FPT algorithm for FVS in mixed graphs. Our algorithm has three main components: The frame of the algorithm is a standard iterative compression approach described in Section 6. The core of our algorithm consists of two parts: the first is a reduction from a variant of FVS to a multi-cut problem called SKEW SEPARATOR. This reduction, described in Section 4 is a non-trivial modification of the reduction employed for FVS in directed graphs by Chen et al [6]. Our reduction only works on pre-conditioned instances, we describe how to perform the necessary pre-conditioning in Section 5.

## 2 Preliminaries

We consider edge/arc labeled multi-graphs: formally, mixed graphs consist of a tuple $G = (V, E, A, \psi)$, where $V$ is the vertex set, $E$ is the edge set, and $A$ is the arc set. The incidence function $\psi$ maps edges $e \in E$ to sets $\{u, v\}$ with $u, v \in V$, also denoted as $uv = vu$. Arcs $a \in A$ are mapped by $\psi$ to tuples $(u, v)$ with $u, v \in V$. In the remainder, we will often denote mixed graphs simply by the tuple $G = (V, E, A)$, and denote $e = uv$ for edges $e \in E$ with $\psi(e) = \{u, v\}$, and $a = (u, v)$ for arcs $a \in A$ with $\psi(a) = (u, v)$. Mixed graphs with $A = \emptyset$ will also denoted by $G = (V, E)$. $V(G)$, $E(G)$ and $A(G)$ denote the vertex, edge and arc set respectively of the mixed graph $G$.

The operation of *contracting* an edge $e = uv$ into a new vertex $w$ consists of the following operations: introduce a new vertex $w$, for every edge or arc with $u$ or $v$ as end vertex, replace this end vertex by $w$, and delete $u$ and $v$. Note that edge identities are preserved: $\psi(e)$ may for instance change from $\{x, u\}$ to $\{x, w\}$, but $e$ is still considered the same edge. Note also that contractions may introduce *parallel edges or arcs* (pairs of edges or arcs $e$ and $f$ with $\psi(e) = \psi(f)$), and *loops* (edges $e$ with $\psi(e) = \{w, w\}$ or arcs $a$ with $\psi(a) = (w, w)$). *Suppressing* a vertex $u$ that is incident with two edges $uv$ and $uw$ ($v = w$ is allowed) means deleting $u$ (and the two incident edges) and adding an edge between $v$ and $w$. We also allow this operation to introduce parallel edges and loops.

For $G = (V, E, A)$ and $S \subseteq V$ or $S \subseteq E \cup A$, by $G[S]$ we denote the subgraph induced by $S$. In particular, $G[E]$ is obtained by deleting all arcs and resulting isolated vertices. Deletion of $S$ is denoted by $G - S$. The *out-degree* $d^+(v)$ (*in-degree* $d^-(v)$) of a vertex $v \in V$ is the number of arcs $e \in A$ with $\psi(e) = (v, w)$ ($\psi(e) = (w, v)$) for some $w$. If an arc $(v, w)$ (($w, v$)) exists, $w$ is called an *out-neighbor* (*in-neighbor*) of $v$. Similarly, the *edge degree* $d(v)$ is the number of incident edges, and if $vw \in E$ then $w$ is an *edge neighbor* of $v$.

A *walk of length* $l$ in a mixed graph $G = (V, E, A)$ is a sequence $v_0, e_1, v_1, e_2, \ldots, e_l, v_l$ such that for all $1 \leq i \leq l$, $e_i \in E \cup A$ and $e_i = v_{i-1}v_i$ or $e_i = (v_{i-1}, v_i)$. This is also called a $(v_0, v_l)$-*walk*. $v_0, v_l$ are its *end vertices*, $v_1, \ldots, v_{l-1}$ its *internal vertices*. A walk is a *path* if all of its vertices are distinct. A walk $v_0, e_1, v_1, \ldots, v_l$ of length at least 1 is a *cycle* if the vertices $v_0, \ldots, v_{l-1}$ are distinct, $v_0 = v_l$, and all $e_i$ are distinct. (Note that this last condition is only relevant for walks of length 2. Note also that if $e$ is a loop on vertex $u$, then $u, e, u$ is also considered a cycle.) We will usually denote walks, paths and cycles just by their vertex sequence $v_0, \ldots, v_l$. In addition, we will sometimes encode paths and cycles by their edge/arc set $E_P = \{e_1, \ldots, e_l\}$.

## 3 Outline of the algorithm

In this section we give an informal overview of our algorithm, the details are given in the following sections. Similar to many previous FVS algorithms [4, 5, 6, 7, 15], we will employ the *iterative compression* technique introduced by Reed, Smith and Vetta [24]. Essentially, this means that we start with a trivial subgraph of $G$ and increase it one vertex at a time until $G$ is obtained, maintaining a FVS of size at most $k + 1$ throughout the computation. Every time we add a vertex to the graph we perform a *compression* step. That is, given a graph $G'$ with a FVS $S$ of size $k + 1$, the algorithm has to decide whether $G'$ has a FVS $S'$ of size $k$. If the algorithm concludes that $G'$ has no FVS of size $k$, we can conclude that $G$ does not either, since $G'$ is a subgraph of $G$. In each compression step the algorithm loops over all $2^{k+1}$ possibilities for $S \cap S'$. For each choice of $S' \cap S$ we need to solve the following problem. See Section 6 for a detailed description of the reduction from FVS to $S$-Disjoint FVS.

$S$-Disjoint FVS:
INSTANCE: A mixed graph $G = (V, E, A)$ with a FVS $S$.
TASK: Find a FVS $S'$ of $G$ with $|S'| < |S|$ and $S' \cap S = \emptyset$, or report that this does not exist.

A FVS $S'$ with $|S'| < |S|$ and $S' \cap S = \emptyset$ is called a *small $S$-disjoint FVS*. Chen et al [6] gave an algorithm for $S$-Disjoint FVS restricted to digraphs, which we will call $S$-*Disjoint Directed FVS*. In Section 4 we show that with some care their algorithm can be extended to solve the following generalization of the problem to mixed graphs. Let $G$ be an undirected graph with $S \subseteq V(G)$. A vertex set $S' \subseteq V(G) \setminus S$ is a *multiway cut* for $S$ (in $G$) if there is no $(u, v)$-path in $G - S'$ for any two distinct $u, v \in S$.

FEEDBACK VERTEX SET / UNDIRECTED MULTIWAY CUT (FVS/UMC):
INSTANCE: A mixed graph $G = (V, E, A)$ with a FVS $S$, and integer $k$.

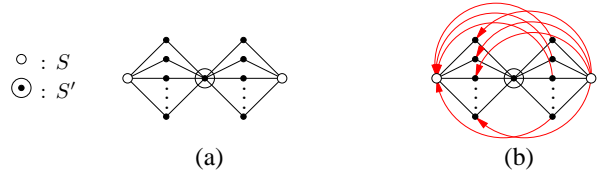

(a)      (b)

Figure 1: Graphs with a FVS $S$ and small $S$-disjoint FVS $S'$, with many undirected $S$-paths.

TASK: Find a FVS $S'$ of $G$ with $|S'| \leq k$ and $S' \cap S = \emptyset$, that is also a multiway cut for $S$ in $G[E]$, or report that this does not exist.

A multiway cut $S'$ for $G[E], S$ is also called an *undirected multiway cut (UMC)* for $G, S$. The remaining question is: how can the FPT algorithm for FVS/UMC be used to solve $S$-Disjoint FVS? Let $G, S$ be an $S$-Disjoint FVS instance. Suppose there exists a small $S$-Disjoint FVS $S'$ for the graph $G$. If we know which *undirected* paths between $S$-vertices do not contain any $S'$-vertices, then these can be contracted, and $S'$ remains a FVS for the resulting graph $G^*$. In addition, this gives a new vertex set $S^*$ consisting of the old $S$-vertices and the vertices introduced by the contractions. This then yields an instance $G^*, S^*$ of FVS/UMC, for which $S'$ is a solution. In Section 5 we prove this more formally. However, since we do not know $S'$, it remains to find which undirected paths between $S$-vertices do not contain $S'$-vertices. One approach would be to try all possible combinations, but the problem is that the number of such paths may not be bounded by any function of $k = |S| - 1$, see the example in Figure 1 (a). (More complex examples with many paths exist, where the solution $S'$ is not immediately obvious.) The example in Figure 1 (a) contains many vertices of degree 2, which are simply reduced in nearly all fast undirected FVS algorithms [7, 25, 15, 4]. However in our case we can easily add arcs to the example to prevent the use of (known) reduction rules, see e.g. Figure 1 (b). Because there may be many such paths, and there are no easy ways to reduce these, we will guess which paths do not contain $S'$-vertices in two stages: this way we only have to consider $2^{O(k)}$ possibilities, which is shown in Section 5.

## 4 An algorithm for FVS/UMC: reduction to Skew Separator

Let $G$ be a digraph and $\mathcal{S} = s_1, \ldots, s_l$ and $\mathcal{T} = t_1, \ldots, t_l$ be mutually disjoint vertex sequences such that all $s_i \in V(G)$ have in-degree 0 and all $t_i \in V(G)$ have out-degree 0. A subset $C \subseteq V(G)$ disjoint from $\{s_1, \ldots, s_l, t_1, \ldots, t_l\}$ is called a *skew separator* if for all $i \geq j$, there is no $(s_i, t_j)$-path in $G - C$. An FPT algorithm to solve the SKEW SEPARATOR problem defined below is given as a subroutine in the algorithm for DIRECTED FEEDBACK VERTEX SET by Chen et al [6].

SKEW SEPARATOR (SS):
INSTANCE: A digraph $G$, vertex sequences $\mathcal{S} = s_1, \ldots, s_l$ and $\mathcal{T} = t_1, \ldots, t_l$ where all $s_i \in V(G)$ have in-degree 0 and all $t_i \in V(G)$ have out-degree 0, and an integer $k$.
TASK: Find a skew separator $C$ of size at most $k$, or report that this does not exist.

**Theorem 1 (Chen et al [6])** *The Skew Separator problem on instances $G, \mathcal{S}, \mathcal{T}, k$ with $n = |V(G)|$ can be solved in time $4^k k \cdot O(n^3)$.*

We will use this to give an algorithm for FVS/UMC, by extending the way SS is used in [6] to give an algorithm for $S$-Disjoint Directed FVS. Let $G$, $S$, $k$ be an instance of FVS/UMC, with $|S| = l$. We define the relation $\prec$ on $V(G)\setminus S$ as follows: $u \prec v$ if and only if there is a $(v,u)$-path in $G - S$ but no $(u,v)$-path (and $u \neq v$). Observe that $\prec$ is transitive and antisymmetric, and therefore a partial order on $V(G)\setminus S$. A bijective function $\sigma : \{1,\ldots,l\} \to S$ is called a *numbering* of $S$. It is an *arc-compatible numbering* if there are no arcs $(\sigma(i), \sigma(j))$ in $G$ with $i > j$.

**Construction** For any numbering $\sigma$ of $S$, the graph $G_{\mathrm{SS}}(G,\sigma)$ is obtained from $G$ as follows: For every $i \in \{1,\ldots,l\}$: denote $v = \sigma(i)$. let $vw_1,\ldots,vw_d$ be the edges incident with $v$, ordered such that if $w_x \prec w_y$ then $x < y$. Since $\prec$ is a partial order, such an ordering exists and is given by an arbitrary linear extension of $\prec$. Apply the following operations:

1. Add the vertices $s_i^1,\ldots,s_i^{d+1}$ and $t_i^1,\ldots,t_i^{d+1}$.

2. For every arc $(v,u)$ with $u \notin S$, add an arc $(s_i^{d+1}, u)$.

3. For every arc $(u,v)$ with $u \notin S$, add an arc $(u, t_i^1)$.

4. For every edge $vw_j$, add arcs $(s_i^j, w_j)$ and $(w_j, t_i^{j+1})$.

5. Delete $v$.

After this is done for every $v \in S$, replace all remaining edges $xy$ with two arcs $(x,y)$ and $(y,x)$. This yields the digraph $G_{\mathrm{SS}}(G,\sigma)$ and vertex sequences $\mathcal{S} = s_1^1,\ldots,s_1^{d_1+1}, s_2^1,\ldots,s_2^{d_2+1}$, $\ldots\ldots, s_l^{d_l+1}$ and $\mathcal{T} = t_1^1,\ldots,t_1^{d_1+1}, t_2^1,\ldots,t_2^{d_2+1},\ldots\ldots,t_l^{d_l+1}$, where $d_i = d(\sigma(i))$ is the edge degree of $\sigma(i)$. The integer $k$ remains unchanged. $G_{\mathrm{SS}}(G,\sigma), \mathcal{S}, \mathcal{T}, k$ is an instance for SS. The following propositions are needed before we can prove a lemma on the relationship between $G, S$ and $G_{\mathrm{SS}}(G,\sigma)$.

**Proposition 2** *Let $G$ be an acyclic mixed graph. If $G$ contains a $(u,v)$-path $P_{uv}$ and a $(v,u)$-path $P_{vu}$, then $P_{uv}$ is an undirected path.*

**Proof:** Suppose that $P_{uv}$ contains at least one arc. Let $P_{uv} = v_0, e_1, v_1, e_2, \ldots, e_{l-1}, v_l$, with all $v_i \in V$ and all $e_i \in E \cup A$.

By induction one can show that if for some $i$, $e_1,\ldots,e_{i-1}$ are also part of the path $P_{vu}$, then these are all edges, and $P_{vu}$ ends with the sub path $v_{i-1}, e_{i-1},\ldots,v_1, e_1, v_0$. Therefore, since $P_{uv}$ contains at least one arc, we can define $i$ to be the smallest index such that $e_i$ is not part of the path $P_{vu}$. Let $j$ be the smallest index $j \geq i$ such that $P_{vu}$ contains the vertex $v_j$ (clearly such a $j$ exists). Since $P_{uv}$ is a path, and $P_{vu}$ ends with the the sub path $v_{i-1}, e_{i-1},\ldots,v_1, e_1, v_0$, it follows that $v_j$ appears before $v_{i-1}$ in the sequence $P_{vu}$. So we can consider the sub path of $P_{uv}$ from $v_{i-1}$ to $v_j$, and the sub path of $P_{vu}$ from $v_j$ to $v_{i-1}$. These paths only share the vertices $v_{i-1}$ and $v_j$, so if one of them has length at least 2, combining them would yield a cycle in $G$. If both have length 1, then combining them yields the walk $v_{i-1}, e_i, v_i, f, v_{i-1}$, for some $f \in E \cup A$. By choice of $e_i$, we have $e_i \neq f$, so this is again a cycle in $G$, a contradiction. $\square$

**Proposition 3** *Let $C \subseteq V(G)\setminus S$ be a FVS and UMC for the graph $G = (V, E, A)$ and vertex set $S \subseteq V$. Then a numbering $\sigma$ of $S$ exists such that for all $1 \leq j < i \leq |S|$, there is no path from $\sigma(i)$ to $\sigma(j)$ in $G - C$.*

**Proof:** Suppose that for some pair $u, v \in S$, $G - C$ contains both a $(u, v)$-path $P_{uv}$ and a $(v, u)$-path $P_{vu}$. Then Proposition 2 shows that $P_{uv}$ is undirected, which contradicts that $C$ is a UMC for $G$. So we can define the following relation $A_R$ on $S$: $(u, v) \in A_R$ if and only if a $(u, v)$-walk exists in $G - C$. By the above argument, the digraph $(S, A_R)$ is acyclic (it is in fact a partial order), so a numbering $\sigma$ of $S$ exists with the desired properties: This is given by a topological ordering of the acyclic digraph $(S, A_R)$ / a linear extension of the partial order $(S, A_R)$. □

**Lemma 4** *Let $S$ be a FVS for a mixed graph $G = (V, E, A)$, such that $G[S]$ contains no edges and $G$ contains no cycles of length less than 3. Then $C \subseteq V(G) \backslash S$ is a FVS and UMC for $G$ and $S$ if and only if there exists an arc-compatible numbering $\sigma$ of $S$ such that $C$ is a skew separator for $G_{SS}(G, \sigma), \mathcal{S}, \mathcal{T}$, as constructed above.*

**Proof:** Let $C$ be a FVS and UMC for $G, S$. By Proposition 3, we can define a numbering $\sigma$ of $S$ such that for all $i > j$, there is no path from $\sigma(i)$ to $\sigma(j)$ in $G - C$. Therefore, $\sigma$ is arc-compatible.

We now show that for this $\sigma$, $C$ is a skew separator for $G_{SS}(G, \sigma), \mathcal{S}, \mathcal{T}$. Let $G_{SS} = G_{SS}(G, \sigma)$. Suppose $C$ is not a skew separator, so $G_{SS} - C$ contains a path $P = s_i^x, v_1, \ldots, v_l, t_j^y$ with $i > j$, or with $i = j$ and $x \geq y$. Then $P' = \sigma(i), v_1, \ldots, v_l, \sigma(j)$ is (the vertex sequence of) a walk in $G - C$; note that arcs of $P$ may correspond to edges in $P'$ but that the vertex sequence still constitutes a walk. If $i > j$, then all vertices of the walk $P'$ are different and hence it is a $(\sigma(i), \sigma(j))$-path in $G - C$, contradicting the choice of $\sigma$. If $i = j$, then $P'$ is a closed walk in $G - C$ of which all internal vertices are distinct. If $P'$ has length at least 3, then all edges/arcs of $P'$ are distinct, so it is a cycle, again a contradiction. If $P'$ has length 1, there is a loop incident with $\sigma(i)$, contradicting the assumption that there are no cycles of length less than 3. Finally suppose the walk $P'$ has length 2, so $P = s_i^x, v_1, t_i^y$. Since $x \geq y$, by the construction of $G_{SS}$, this can only occur when $x = d+1$ or $y = 1$. Therefore $G$ contains an arc $(\sigma(i), v_1)$ or an arc $(v_1, \sigma(i))$. It follows that distinct arcs/edges $e$ and $f$ can be chosen such that $P' = \sigma(i), e, v_1, f, \sigma(i)$ is a cycle of length 2 in $G$, again a contradiction. Therefore, $C$ is a skew separator for $G_{SS}$.

Let $C$ be a skew separator for $G_{SS} = G_{SS}(G, \sigma)$, for some arc-compatible numbering $\sigma$ of $S$. We prove that $C$ is a FVS and UMC for $G, S$. Suppose $G[E] - C$ contains a $(u, v)$-path $P = u, v_1, \ldots, v_l, v$ with distinct $u, v \in S$, and no internal vertices in $S$. Let $u = \sigma(i)$ and $v = \sigma(j)$. Since we assumed that $G[S]$ contains no edges, $P$ has length at least 2. Since all edges not incident with $S$ are replaced with arcs in both directions during the construction of $G_{SS}$, for some $x, y$ this yields both a path $s_i^x, v_1, \ldots, v_l, t_j^{y+1}$ in $G_{SS} - C$ and a path $s_j^y, v_l, \ldots, v_1, t_i^{x+1}$ in $G_{SS} - C$. One of these paths contradicts that $C$ is a skew separator (depending on whether $i < j$ or $j < i$). This shows that $C$ is a multiway cut for $G[E]$ and $S$.

Next, suppose $G - C$ contains a cycle $K$. Since $S$ is a FVS for $G$, $K$ contains at least one vertex of $S$. If $K$ contains at least two vertices of $S$, then $K$ contains a path $P$ from $\sigma(i)$ to $\sigma(j)$ for some $i > j$, with no internal vertices in $S$. Let $P = \sigma(i), v_1, \ldots, v_l, \sigma(j)$. $P$ has length at least two, since $\sigma$ is arc-compatible, and there are no edges in $G[S]$. Then $P' = s_i^x, v_1, \ldots, v_l, t_j^y$ is a path in $G_{SS} - C$ for some $x, y$, contradicting that $C$ is a skew separator.

So now we may suppose that $K$ contains exactly one vertex of $S$, w.l.o.g. $K = \sigma(i), v_1, \ldots, v_l, \sigma(i)$. Every cycle in $G$ has length at least 3, so $v_1 \neq v_l$. In the case that $(\sigma(i), v_1) \in A$, $K$ yields a path $P = s_i^{d+1}, v_1, \ldots, v_l, t_i^y$ in $G_{\text{SS}} - C$ for some $y \leq d+1$, a contradiction (here $d = d(\sigma(i))$ is the edge degree of $\sigma(i)$). On the other hand, if $(v_l, \sigma(i)) \in A$, then $K$ yields a path $P = s_i^x, v_1, \ldots, v_l, t_i^1$ in $G_{\text{SS}} - C$ for some $x \geq 1$, a contradiction. So finally suppose that $\sigma(i)v_1 \in E$ and $\sigma(i)v_l \in E$ are both edges. Then $K$ gives a path $s_i^x, v_1, \ldots, v_l, t_i^{y+1}$ in $G_{\text{SS}} - C$ for some $x, y$. Since $C$ is a skew separator, $x \leq y$. Since $v_1 \neq v_l$, $x < y$. Therefore $v_l \not\prec v_1$. The cycle $K$ shows that there is a $(v_1, v_l)$-path $P$ in $G - S$. Then, by the definition of $\prec$, there must also be a $(v_l, v_1)$-path in $G - S$. But this can only happen if $P$ is an undirected path (Proposition 2). This shows that by reversing the cycle $K$, we again obtain a cycle $\sigma(i), v_l, v_{l-1}, \ldots, v_1, \sigma(i)$ in $G - C$, and therefore a path $s_i^y, v_l, v_{l-1}, \ldots, v_1, t_i^{x+1}$ in $G_{\text{SS}} - C$, a contradiction (since $x < y$). This concludes all cases, so $C$ is a FVS for $G$.

This concludes the proof that $C$ is a FVS and UMC for $G, S$. □

**Theorem 5** *FVS/UMC on instances $G, S, k$ with $n = |V(G)|$, $k \geq 1$ and $l = |S|$ can be solved in time $O(n^3) \cdot l! \, 4^k k$.*

**Proof:** We may return 'NO' immediately if $G[S]$ contains edges, or if $G[S]$ contains cycles. The latter holds in particular if $G$ contains loops. So suppose none of this holds. Then if $G$ contains a cycle $C$ of length 2, $C$ must contain one $S$-vertex and one non-$S$-vertex $v$. Every FVS/UMC solution contains $v$, so we may reduce the instance by deleting $v$ and decreasing $k$ by one, to obtain an equivalent instance. Furthermore, if any vertex $u \notin S$ has an edge to at least two distinct vertices in $S$ then any undirected multiway cut for $S$ must contain $u$. Hence we may reduce the instance by deleting $u$ and decreasing $k$ by one. So we may now assume w.l.o.g. that $G$ contains no cycles of length less than 3 and no two vertices in $S$ have edges to the same vertex in $V(G) \setminus S$. To find a FVS and UMC, we try all arc-compatible numberings $\sigma$ of $S$, and test whether $G_{\text{SS}}(G, \sigma)$ has a skew separator of size at most $k$. There are at most $l!$ such numberings. Return such a skew separator $C$ if it is found for any arc-compatible numbering $\sigma$, or 'NO' otherwise. By Lemma 4, this correctly solves FVS/UMC. Note that $G_{\text{SS}}(G, \sigma)$ can be constructed in time $O(n^3)$. Since no two vertices in $S$ have edges to the same vertex in $V(G) \setminus S$, $G_{\text{SS}}(G, \sigma)$ has at most $3n$ vertices. Thus, for every choice of $\sigma$, the complexity is bounded by $O(n^3) \cdot 4^k k$ (Theorem 1). □

## 5  An algorithm for $S$-Disjoint FVS: contracting paths

In this section we give an FPT algorithm for $S$-Disjoint FVS, by reducing it to FVS/UMC. Throughout this section, let $G = (V, E, A)$ be a mixed graph, and $S$ be a FVS for $G$. For a given mixed graph $G = (V, E, A)$ and FVS $S$ we define an undirected graph $\text{BB}(G, S)$ (short for *backbone*) together with a mapping $P_G(G, S)$ from the edges of $\text{BB}(G, S)$ to edge sets of $G$ as follows. See also Figure 2. Start with $G[E]$, and for all $e \in E$, set $P_G(e) = \{e\}$. First, iteratively delete all vertices of (edge) degree at most 1 that are not in $S$. Next, iteratively suppress all degree 2 vertices that are not in $S$, until a graph is obtained in which every non-$S$-vertex has degree at least 3. (Recall that suppressing vertices may yield multi-edges.) Whenever a degree two vertex $u$ with neighbors $v$ and $w$ is suppressed, we set $P_G(vw) = P_G(uv) \cup P_G(uw)$ for the new edge $vw$. This yields the undirected multi-graph $\text{BB}(G, S)$, and mapping $P_G(G, S)$.

Let $\text{BB} = \text{BB}(G, S)$ and $P_G = P_G(G, S)$. Since during the construction of $\text{BB}$, only non-$S$-vertices are suppressed, the property that every cycle contains an $S$-vertex is maintained, so $S$ is a FVS for $\text{BB}(G,S)$. Non-$S$-vertices of $G$ that are part of BB are called the *branching vertices* of $G$ (with respect to $S$), notated by $\text{Br}(G,S)$. Let $\text{Br} = \text{Br}(G,S)$.

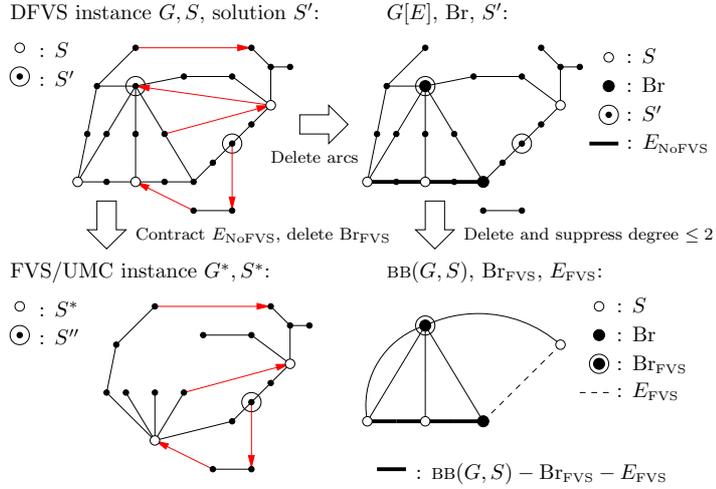

Figure 2: The graphs and sets defined in Section 5.

**Proposition 6** *Let $\text{BB} = \text{BB}(G,S)$, for a mixed graph $G = (V, E, A)$. There exists a path (cycle) $P = v_0, e_1, v_2, \ldots, e_l, v_l$ in $G[E]$ with $v_0, v_l \in S \cup \text{Br}$, and no other vertices in $S \cup \text{Br}$, if and only if there is a non-loop edge $e = v_0 v_l \in E(\text{BB})$ (resp. a loop $e \in E(\text{BB})$ on $v_0 = v_l$), with $P_G(e) = \{e_1, \ldots, e_l\}$.*

**Proof:** Consider the case where $P$ is a path. The cycle case is similar. During the construction of BB, none of the internal vertices of $P$ will receive degree 1. So since they are not in $S \cup \text{Br} = V(\text{BB})$, they are all suppressed. Regardless of the order in which this happens, this results in an edge $e = v_0 v_l \in E(\text{BB})$ with $P_G(e) = \{e_1, \ldots, e_l\}$.

Now we prove the other direction. A simple induction argument shows that at any time during the construction of BB, for all (non-loop) edges $e$, the edge set $P_G(e)$ induces a path $P$ in $G$, where all internal vertices of $P$ have been suppressed earlier. The statement follows. □

The paths $P$ in $G$ as in Proposition 6 (undirected paths with end vertices in $S \cup \text{Br}$ and no internal vertices in $S \cup \text{Br}$) are called *connection paths* of $G$ (with respect to $S$). Proposition 6 will be used often (implicitly) in the proofs below, for instance as follows. Let $S' \subseteq V \backslash S$. Define $\text{Br}_{\text{FVS}} = \text{Br} \cap S'$, and define $E_{\text{FVS}} \subseteq E(\text{BB})$ to be those edges $uv \in E(\text{BB})$ with $u, v \notin S'$, but where some edge in $P_G(uv)$ is incident with $S'$. Then we refer to the above proposition to show that $G[E] - S'$ contains a cycle if and only if $\text{BB} - \text{Br}_{\text{FVS}} - E_{\text{FVS}}$ contains a cycle. (Observe also that two connection paths can not share internal vertices.)

**Lemma 7** *Let $S$ be a FVS of a mixed graph $G = (V, E, A)$ with $k = |S| - 1$, and let $S'$ be a small $S$-disjoint FVS for $G, S$. Then $G$ has at most $3k$ branching vertices with respect to $S$, and $G$ has at most $3k$ connection paths with no vertices in $S'$.*

**Proof:** Let $\text{Br} = \text{Br}(G,S)$, $\text{BB} = \text{BB}(G,S)$, and $P_G = P_G(G,S)$. Denote by $E_{\text{FVS}} \subseteq E(\text{BB})$ the set of edges $e$ for which the path $P_G(e)$ in $G[E]$ has an internal vertex in $S'$, but no end vertices in $S'$. Let $\text{Br}_{\text{FVS}} = \text{Br} \cap S'$.

Recall that BB − S contains no cycles. Therefore, it is possible to orient all edges of BB such that every non-S-vertex has at most one in-neighbor (for every tree in the forest BB − S, choose an arbitrary root, and orient all edges away from the root. Orient all other edges towards S-vertices). In the rest of the proof, out-degrees, denoted by $d^+(v)$, will refer to such an orientation of BB. Now, $|V(\text{BB})| = |\text{Br}| + |S|$. Let $G_U = (V_U, E_U) = \text{BB} - \text{Br}_{\text{FVS}} - E_{\text{FVS}}$. Since $S'$ is a FVS of $G$, $G_U$ contains no cycles (Proposition 6), so

$$|E_U| \leq |V_U| - 1 = |\text{Br}| + |S| - |\text{Br}_{\text{FVS}}| - 1. \tag{1}$$

Since vertices in Br have out-degree at least 2 we obtain: $2|\text{Br}\setminus\text{Br}_{\text{FVS}}| - |\text{Br}_{\text{FVS}}| - |E_{\text{FVS}}| \leq \sum_{v \in \text{Br}\setminus\text{Br}_{\text{FVS}}} d^+(v) - |\text{Br}_{\text{FVS}}| - |E_{\text{FVS}}| \leq |E_U|$. The last inequality holds since every non-$S$ vertex, in particular every $\text{Br}_{\text{FVS}}$-vertex, has in-degree at most 1. So deleting a $\text{Br}_{\text{FVS}}$-vertex removes at most one arc that counts towards the out-degree of another vertex. Substituting Inequality 1 for $|E_U|$ then yields:

$$|\text{Br}| \leq 2|\text{Br}_{\text{FVS}}| + |E_{\text{FVS}}| + |S| - 1. \tag{2}$$

Since $|\text{Br}_{\text{FVS}}| + |E_{\text{FVS}}| \leq |S'|$, the above inequality immediately yields $|\text{Br}| \leq 2|S'| + |S| - 1 \leq 3k$, proving the first statement of the lemma. Secondly, combining Inequality 1 with Inequality 2 yields that $|E_U| \leq |\text{Br}| + |S| - |\text{Br}_{\text{FVS}}| - 1 \leq |\text{Br}_{\text{FVS}}| + |E_{\text{FVS}}| + 2|S| - 2 \leq |S'| + 2|S| - 2 \leq 3k$. Since there are $|E_U|$ connection paths without $S'$-vertices, this concludes the proof. □

Algorithm 1 shows the algorithm for solving $S$-Disjoint FVS. Recall that the 'continue' statement continues with the next iteration of the smallest enclosing for- (or while-) loop, so it skips the remainder of the current iteration.

**Algorithm 1** An algorithm for $S$-Disjoint FVS

INPUT: A mixed graph $G = (V, E, A)$ with FVS $S$, and integer $k = |S| - 1$.
OUTPUT: a small $S$-disjoint FVS $S'$ for $G, S$, or 'NO' if this does not exist.
1.    BB := BB$(G, S)$, $P_G := P_G(G, S)$ and Br := Br$(G, S)$.
2.    **if** $|\text{Br}| > 3k$ **then return** 'NO'
3.    **for** all $\text{Br}_{\text{FVS}} \subseteq \text{Br}$ with $|\text{Br}_{\text{FVS}}| \leq k$:
4.        $k' := k - |\text{Br}_{\text{FVS}}|$.
5.        $G' := \text{BB}(G, S) - \text{Br}_{\text{FVS}}$.
6.        **if** $|E(G')| > 3k + k'$ **then continue**
7.        **for** all $E_{\text{FVS}} \subseteq E(G')$ with $|E_{\text{FVS}}| \leq k'$:
8.            **if** $G' - E_{\text{FVS}}$ contains a cycle **then continue**
9.            $E_{\text{NoFVS}} := \cup_{e \in E(G')\setminus E_{\text{FVS}}} P_G(e)$
10.           $G^* := G - \text{Br}_{\text{FVS}}$, $S^* := S$.
11.           **while** there is an edge $uv \in E(G^*) \cap E_{\text{NoFVS}}$ with $v \in S^*$:
12.               Change $G^*$ by contracting $e$ into a new vertex $w$
13.               $S^* := (S^*\setminus\{u, v\}) \cup \{w\}$
14.           **if** $G^*$ contains no loops incident with $S^*$-vertices and
              there is a FVS and UMC $S''$ for $G^*, S^*$ with $|S''| \leq k'$, **then**
15.               **return** $S' := S'' \cup \text{Br}_{\text{FVS}}$
16.   **return** 'NO'

Before analyzing the complexity of Algorithm 1, we first give variants of often used bounds on sums of binomial coefficients. Proofs are included for completeness.

**Proposition 8** *For all $c > 2$, $\sum_{i=0}^{k} \binom{ck}{i} < \frac{c-1}{c-2}\binom{ck}{k}$.*

**Proof:** For all $i \leq k$,

$$\binom{ck}{i-1} \bigg/ \binom{ck}{i} = \frac{i!(ck-i)!}{(i-1)!(ck-i+1)!} = \frac{i}{ck-i+1} \leq \frac{i}{(c-1)i+1} < \frac{1}{c-1},$$

So we may write

$$\sum_{i=0}^{k} \binom{ck}{i} < \binom{ck}{k} \sum_{i=0}^{k} \left(\frac{1}{c-1}\right)^{k-i} < \frac{1}{1-\frac{1}{c-1}} \binom{ck}{k} = \frac{c-1}{c-2}\binom{ck}{k}.$$

□

**Proposition 9** *For all constants $c > 1$, $\binom{ck}{k} \in O\left(\left(\frac{c^c}{(c-1)^{c-1}}\right)^k\right)$.*

**Proof:** By Stirling's approximation $n! \in \Theta\left(n^n e^{-n} \sqrt{n}\right)$,

$$\binom{ck}{k} \in O\left(\frac{(ck)^{ck} e^{-ck} \sqrt{ck}}{((c-1)k)^{(c-1)k} e^{-(c-1)k} \sqrt{(c-1)k} \; k^k e^{-k} \sqrt{k}}\right) \subset$$

$$O\left(\frac{(ck)^{ck}}{((c-1)k)^{(c-1)k} k^k}\right) = O\left(\left(\frac{c^c}{(c-1)^{c-1}}\right)^k\right).$$

□

**Theorem 10** *On an instance $G = (V, E, A), S$ with $n = |V|$ and $k = |S| - 1$, Algorithm 1 correctly solves S-Disjoint FVS in time $O\left(k(k+1)! \; 47.5^k \; n^3\right)$.*

**Proof:** We first show that the correct answer is returned. Let $\text{BB} = \text{BB}(G, S)$, $P_G = P_G(G, S)$ and $\text{Br} = \text{Br}(G, S)$. Consider the case that a set $S' = S'' \cup \text{Br}_{\text{FVS}}$ is returned in Line 15. Let $\text{Br}_{\text{FVS}}$, $E_{\text{FVS}}$ and $G' = \text{BB} - \text{Br}_{\text{FVS}}$ refer to these sets/graph as they are in the corresponding iteration of the for-loops. Clearly, $|S'| \leq k' + |\text{Br}_{\text{FVS}}| = k$, and $S' \cap S = \emptyset$. So to show that $S'$ is a small S-disjoint FVS of $G$, it only remains to verify that it is a FVS for $G$. Let $E_{\text{CONTR}}$ be the edges of $G$ that are contracted to obtain $G^*$ (in Line 12), so $E_{\text{CONTR}} \subseteq E_{\text{NoFVS}}$, with $E_{\text{NoFVS}} = \cup_{e \in E(G') \setminus E_{\text{FVS}}} P_G(e)$ (Line 9). All edges of $E_{\text{CONTR}}$ are part of connection paths. Suppose $G - S'$ contains a cycle $C$, encoded by its edge and arc set (so $C \subseteq E \cup A$). If $C$ consists only of edges that are in $E_{\text{CONTR}}$, then $G' - E_{\text{FVS}}$ contains a cycle as well (Proposition 6). But then the algorithm would have continued to the next for-loop iteration in Line 8, a contradiction. So contracting all edges of $C$ that are in $E_{\text{CONTR}}$ does not contract $C$ into a single vertex, but yields a cycle $C'$ in $G^*$: note that $C$ contains no vertices of $\text{Br}_{\text{FVS}}$ so it does not matter that these vertices are deleted for the construction of $G^*$. Since $C'$ also contains no vertices of $S'' \subseteq S'$, this contradicts that $S''$ is a FVS of $G^*$. This concludes the correctness proof in case a positive answer is returned.

Now suppose a small S-disjoint FVS $S'$ exists for $G$. We will show that then a positive answer is returned (in Line 15). First, by Lemma 7, $|\text{Br}| \leq 3k$, so Line 2 does not terminate by returning 'NO'. Let $\text{Br}_{\text{FVS}} = S' \cap \text{Br}$. Consider the iteration of the first for-loop where $\text{Br}_{\text{FVS}}$ is considered. Let $E_{\text{FVS}}$ be the edges of $\text{BB}(G, S)$ that correspond to connection paths with no end vertices in $S'$ but at least one internal vertex in $S'$. Then $|E_{\text{FVS}}| + |\text{Br}_{\text{FVS}}| \leq |S'| \leq k$, so $|E_{\text{FVS}}| \leq k' = k - |\text{Br}_{\text{FVS}}|$. Let $G' = \text{BB} - \text{Br}_{\text{FVS}}$. All edges $e \in E(G')\backslash E_{\text{FVS}}$ correspond to connection paths $P_G(e)$ with no vertices in $S'$, so by Lemma 7, $|E(G')\backslash E_{\text{FVS}}| \leq 3k$, and therefore $|E(G')| \leq 3k + k'$. This shows that in Line 6 the algorithm will not continue to the next for-loop iteration, and that therefore $E_{\text{FVS}}$ will be considered in the second for-loop. A cycle in $G' - E_{\text{FVS}}$ would correspond to a cycle in $G - S'$ (Proposition 6), so the algorithm also does not skip the remainder of the for-loop iteration in Line 8. Therefore, in Line 10–13, the graph $G^*$ and vertex set $S^*$ is constructed from $G$ by deleting $\text{Br}_{\text{FVS}}$ and contracting edges in $E_{\text{NoFVS}}$, which are part of connection paths and not incident with vertices in $S'$. We will now argue that $S'' = S'\backslash \text{Br}_{\text{FVS}}$ is a FVS and UMC for $G^*$ and $S^*$, so a positive answer is returned in Line 15. (Clearly, $|S''| \leq k - |\text{Br}_{\text{FVS}}| = k'$.)

Firstly, $S'$ is a FVS for $G$, so $S'' = S'\backslash \text{Br}_{\text{FVS}}$ is a FVS for $G - \text{Br}_{\text{FVS}}$. This shows that when $G^*$ is constructed in Line 10, $S''$ is a FVS for it. Contracting (undirected!) edges that are not incident with $S''$-vertices maintains this property, so after the while-loop in Line 11 has terminated, $S''$ is still a FVS for $G^*$.

Secondly, we show that $S''$ is a UMC for $G^*$. We prove that throughout the while-loop (Line 11), the following invariant holds:

*Property 1:* For every undirected path $P \subseteq E(G^*)$ with end vertices in $S^*$ and no vertices in $S''$, $P \subseteq E_{\text{NoFVS}}$.

Every undirected path $P$ between two $S$-vertices in $G$ consists of a sequence $P^1, \ldots, P^l$ of connection paths. If $P$ is a path in $G - \text{Br}_{\text{FVS}}$, then all of these paths $P^i$ correspond to edges of the graph $G' = \text{BB} - \text{Br}_{\text{FVS}}$ as constructed in Line 5, i.e. $P^i = P_G(e)$ for some $e \in E(G')$. If $P$ is not incident with vertices in $S'' = S'\backslash \text{Br}_{\text{FVS}}$, then these paths $P^i$ correspond to edges $e \in E(G')\backslash E_{\text{FVS}}$. This proves that $P \subseteq E_{\text{NoFVS}} = \cup_{e \in E(G')\backslash E_{\text{FVS}}} P_G(e)$, so initially the property holds. Next, observe that contracting an edge $e \in E_{\text{NoFVS}}$ incident with an $S^*$-vertex but not with an $S''$ vertex, and adding the resulting vertex to $S^*$ again (Line 13), does not destroy the property. We conclude that, after the while-loop (Line 11) has terminated, there are no undirected paths in $G^*$ between two $S^*$-vertices that contain no $S''$-vertices. This shows that $S''$ is a UMC for the resulting $G^*$ and $S^*$.

We conclude that if a small $S$-disjoint FVS $S'$ exists, then in at least one of the iterations of the for-loops, a positive answer will be returned in Line 15. This concludes the correctness proof of the algorithm.

Now we will consider the complexity of the algorithm. First, consider the parameter function. By Line 2, $|\text{Br}| \leq 3k$, so the number of iterations of the first for-loop is at most

$$\sum_{i=0}^{k} \binom{3k}{i}.$$

Here $i = |\text{Br}_{\text{FVS}}|$. Let $k' = k - i$. By Line 6, $|E(G')| \leq 3k + k'$ holds for $G'$ whenever the second for-loop is entered, so there are at most

$$\sum_{j=0}^{k'} \binom{3k + k'}{j} = \sum_{j=0}^{k-i} \binom{4k - i}{j} < \frac{3}{2} \binom{4k - i}{k - i}$$

choices of $E_{\text{FVS}}$ (Proposition 8). So we may bound the total number of iterations of the second for-loop by

$$\frac{3}{2} \sum_{i=0}^{k} \binom{3k}{i} \binom{4k - i}{k - i}.$$

At most once for every iteration, a FVS/UMC problem on the instance $G^*$, $S^*$, $k'$ is solved, which can be done with parameter function $|S^*|! \cdot 4^{k'} k'$ (Theorem 5). Initially (Line 10), $|S^*| = |S| = k+1$. Throughout the while-loop, the size of $S^*$ does not increase. Therefore the parameter function of Algorithm 1 is bounded by a constant times

$$\sum_{i=0}^{k} \binom{3k}{i} \binom{4k-i}{k-i} (k+1)! 4^{k-i} \max\{1, k-i\} \leq$$

$$k(k+1)! \sum_{i=0}^{k} \frac{(3k)!}{i!(3k-i)!} \frac{(4k-i)!}{(k-i)!(3k)!} 4^{k-i} =$$

$$k(k+1)! \sum_{i=0}^{k} \binom{k}{i} \binom{4k-i}{k} 4^{k-i} <$$

$$k(k+1)! \binom{4k}{k} \sum_{i=0}^{k} \binom{k}{i} 4^{k-i} \in$$

$$O\left(k(k+1)!\ 9.5^k\ (1+4)^k\right) = O\left(k(k+1)!\ 47.5^k\right).$$

For the last line, we used Proposition 9, and $\frac{4^4}{3^3} = \frac{256}{27} < 9.5$.

Now we prove that the polynomial part of the complexity (the complexity for fixed $k$) can be bounded by $O(n^3)$, where $n = |V|$. Let $m = |E| + |A|$. Although we allow multi-graphs, w.l.o.g. we may assume $m \in O(n^2)$. Graphs are encoded with adjacency lists in such a way that edges can be deleted in constant time, vertices $v$ can be deleted in time $O(d^T(v))$, and edges $uv$ can be contracted in time $O(d^T(u) + d^T(v))$, where $d^T(v) = d(v) + d^+(v) + d^-(v)$ denotes the total number of arcs and edges incident with $v$. For most steps in the algorithm (that we did not already attribute to the parameter function) it can now be verified that they can be done in constant time or linear time $O(n+m) \subseteq O(n^2)$. (Lines 1, 5, 8, 9 and 10 require linear time.) Only Line 14 and the while-loop in Line 11 need further consideration. If the while-loop of Line 11 is entered, then $G' - E_{\text{FVS}}$ contains no cycles (Line 8). Therefore $G[E_{\text{NoFVS}}]$ contains no cycles (Proposition 6), and thus the while-loop iterates at most $n-1$ times. Each iteration requires at most time $O(m)$, which gives a complexity of $O(nm) \subseteq O(n^3)$. Evaluating in Line 14 whether a FVS and UMC exists takes time $O(n^3)$ as well for fixed $k$ (Theorem 5). This proves that the total complexity is $O\left(k(k+1)!\ 47.5^k\ n^3\right)$. □

## 6 An algorithm for FVS: iterative compression

We now show how the iterative compression technique gives an algorithm for FVS. This is similar to many previous FVS algorithms [4, 5, 6, 7, 15], but we state the details for completeness. Algorithm 2 shows how to solve the *FVS Compression Problem*: given a FVS of size $k+1$ for a mixed graph $G$, decide if there exists a FVS of size at most $k$, and if so, return one.

**Theorem 11** *If $S$ is a FVS of $G = (V, E, A)$ with $|S| = k+1$, then Algorithm 2 decides in time $O\left((k+1)! k^2 47.5^k\ n^3\right)$ whether $G$ admits a FVS of size at most $k$.*

**Algorithm 2** An algorithm for FVS compression

INPUT: A mixed graph $G = (V, E, A)$ with a FVS $S$, $k = |S| - 1$.
OUTPUT: A FVS $S'$ of $G$ with $|S'| \leq k$, or 'NO' if this does not exist.

1.    **for** all $S_{\text{DEL}} \subseteq S$ with $|S_{\text{DEL}}| \geq 1$:
2.        $S_{\text{KEEP}} := S \backslash S_{\text{DEL}}$, $G' = G - S_{\text{KEEP}}$.
3.        **if** $G'$ has an $S_{\text{DEL}}$-Disjoint FVS $S_{\text{NEW}}$ with $|S_{\text{NEW}}| \leq |S_{\text{DEL}}| - 1$ **then**
4.           **return** $S_{\text{NEW}} \cup S_{\text{KEEP}}$
5.    **return** 'NO'

---

**Proof:** First suppose that a set $S' = S_{\text{NEW}} \cup S_{\text{KEEP}}$ is returned in Line 4. Then $|S'| \leq |S_{\text{DEL}}| - 1 + |S_{\text{KEEP}}| = |S| - 1 = k$. If a cycle $C$ of $G$ contains an $S_{\text{KEEP}}$-vertex, then it contains an $S'$-vertex. Otherwise, it is a cycle of $G'$, and thus it contains a vertex in $S_{\text{NEW}} \subseteq S'$. This shows that $S'$ is a FVS of size at most $k$.

Now suppose that $G$ contains a FVS $S'$ with $|S'| \leq k$. Let $S_{\text{KEEP}} = S' \cap S$, $S_{\text{DEL}} = S \backslash S'$ and $S_{\text{NEW}} = S' \backslash S$. Since $S_{\text{NEW}} = S' \backslash S_{\text{KEEP}}$, $S_{\text{NEW}}$ is a FVS of $G' = G - S_{\text{KEEP}}$, of size at most $|S_{\text{NEW}}| = |S'| - |S_{\text{KEEP}}| \leq k - |S \backslash S_{\text{DEL}}| = k - |S| + |S_{\text{DEL}}| = |S_{\text{DEL}}| - 1$. This shows that in the iteration where $S_{\text{DEL}}$ is considered, a positive answer is returned. This concludes the correctness proof.

Now we consider the complexity. If $|S_{\text{DEL}}| = j + 1$, then deciding the condition in Line 3 takes time $O\left(j(j+1)! \, 47.5^j \, n^3\right)$ (Theorem 10). There are $\sum_{j=0}^{k} \binom{k+1}{j+1} = k + 1 + \sum_{j=1}^{k} \binom{k+1}{j+1}$ choices of $S_{\text{DEL}}$ to consider in the for-loop. This yields a complexity in the order of

$$\sum_{j=1}^{k} \binom{k+1}{j+1} j(j+1)! \, 47.5^j \, n^3 = (k+1)! \, n^3 \sum_{j=1}^{k} \frac{j 47.5^j}{(k-j)!} < (k+1)! k^2 47.5^k \, n^3.$$

□

**Theorem 12** *In time $O\left((k+1)! k^2 47.5^k \, n^4\right)$, it can be decided whether a mixed graph $G = (V, E, A)$ with $|V| = n$ contains a FVS $S$ with $|S| \leq k$.*

**Proof:** Let $V = \{v_1, \ldots, v_n\}$, and for all $1 \leq i \leq n$, let $G_i = G[\{v_1, \ldots, v_i\}]$. Clearly, $G_k$ has a FVS of size at most $k$. In addition, for every $k \leq i \leq n - 1$, if $S$ is a FVS for $G_i$ with $|S| \leq k$, then $S' = S \cup \{v_{i+1}\}$ is a FVS for $G_{i+1}$ with $|S'| \leq k + 1$. This shows that, in order to decide whether $G = G_n$ admits a FVS of size $k$, we only need to solve the FVS Compression Problem at most $n - k$ times, once for every $k + 1 \leq i \leq n$. By Theorem 11, the total complexity then becomes $(n-k) \cdot O\left((k+1)! k^2 47.5^k \, n^3\right) \subseteq O\left((k+1)! k^2 47.5^k \, n^4\right)$. □

*Remark:* By abusing the O-notation, the complexity of our algorithm can also be bounded by $O\left(k! \, 47.5^k \, n^4\right)$: Observe that by rounding less generously in the proof of Theorem 10, the complexity of our FVS algorithm becomes $O\left((k+1)! k^2 47.41^k \, n^4\right)$. Since $k^3 \in O\left((1+\epsilon)^k\right)$ holds for every $\epsilon > 0$, this yields

$$O\left((k+1)! k^2 47.41^k \, n^4\right) \subseteq O\left(k! 47.5^k \, n^4\right).$$